\documentclass[prx,twocolumn,superscriptaddress,longbibliography,floatfix]{revtex4-2}

\usepackage[SquareTraceBrackets]{quantum}
\usepackage{graphicx,bm,natbib,upgreek,amsmath,mathrsfs,accents}
\usepackage{amsbsy}
\usepackage[dvipsnames]{xcolor}
\definecolor{myblue}{named}{MidnightBlue}
\definecolor{mygreen}{RGB}{0,120,0}
\usepackage[colorlinks=true,citecolor=myblue,linkcolor=myblue,urlcolor=myblue]{hyperref}
\usepackage[T1]{fontenc}
\usepackage{newtxtext,newtxmath}

\usepackage[scaled]{helvet}
\usepackage{tikz}
\usetikzlibrary{arrows,decorations.pathmorphing,backgrounds,positioning,fit,petri}
\usepackage{gensymb}				
\usepackage[caption=false]{subfig}
\usepackage{enumitem}
\usepackage{tabularx}
\usepackage{multirow}
\usepackage[english]{isodate,babel}
\usepackage{todonotes}

\makeatletter
\def\thickhline{%
  \noalign{\ifnum0=`}\fi\hrule \@height \thickarrayrulewidth \futurelet
   \reserved@a\@xthickhline}
\def\@xthickhline{\ifx\reserved@a\thickhline
               \vskip\doublerulesep
               \vskip-\thickarrayrulewidth
             \fi
      \ifnum0=`{\fi}}
\makeatother

\usepackage{amsthm}

\DeclareGraphicsExtensions{.pdf, .jpg, .eps, .svg}

\newcommand{\slm}{\eta_\text{loss}^\text{sys}}

\begin{document}

\title{Finite key performance of satellite quantum key distribution under practical constraints}

\author{Jasminder S. Sidhu}
\email{jsmdrsidhu@gmail.com}
\thanks{Corresponding author email}
\affiliation{SUPA Department of Physics, University of Strathclyde, Glasgow, G4 0NG, United Kingdom}
\author{Thomas Brougham}
\affiliation{SUPA Department of Physics, University of Strathclyde, Glasgow, G4 0NG, United Kingdom}
\author{Duncan McArthur}
\affiliation{SUPA Department of Physics, University of Strathclyde, Glasgow, G4 0NG, United Kingdom}
\author{Roberto G. Pousa}
\affiliation{SUPA Department of Physics, University of Strathclyde, Glasgow, G4 0NG, United Kingdom}
\author{Daniel K. L. Oi}
\affiliation{SUPA Department of Physics, University of Strathclyde, Glasgow, G4 0NG, United Kingdom}

\date{\today}

\begin{abstract}
Global-scale quantum communication networks will require efficient long-distance distribution of quantum signals. Optical fibre communication channels have range constraints due to exponential losses in the absence of quantum memories and repeaters. Satellites enable intercontinental quantum communication by exploiting more benign inverse square free-space attenuation and long sight lines. However, the design and engineering of satellite quantum key distribution (QKD) systems is difficult and characteristic differences to terrestrial QKD networks and operations pose additional challenges. The typical approach to modelling satellite QKD (SatQKD) has been to estimate performances with a fully optimised protocol parameter space and with few payload and platform resource limitations. Here, we analyse how practical constraints affect the performance of SatQKD for the Bennett-Brassard 1984 (BB84) weak coherent pulse decoy state protocol with finite key size effects. We consider engineering limitations and trade-offs in mission design including limited in-orbit tunability, quantum random number generation rates and storage, and source intensity uncertainty. We quantify practical SatQKD performance limits to determine the long-term key generation capacity and provide important performance benchmarks to support the design of upcoming missions.
\end{abstract}

\maketitle

\section{Introduction}
\label{sec:intro}

\noindent
Quantum technologies have the potential to enable or greatly enhance applications including secure communications~\cite{QKDreview2020, sidhu2021advances, Liorni2021, Wallnofer2022}, improved computation~\cite{Donkor2004distributed, Meter2016the}, sensing, and imaging~\cite{Giovannetti2011_NP, Sidhu2017_PRA, Sidhu2018_arxiv, Moreau2019_NRP, Sidhu2020_AVS, Polino2020_AVS, Sidhu2021_PRX}. In addition, a distributed ecosystem of quantum technologies would provide further performance improvements and additional capabilities. The distribution of quantum resources across such a networked architecture comprises the fundamental building blocks of the quantum internet~\cite{Wallnofer2022}. 

Satellites will be integral to a scalable architecture to expand the range of quantum networks to global scales, motivating the surge in recent activities in space quantum communications~\cite{Liao2017_N, Yin2017_S,Kerstel2018_EPJ, Mazzarella2020_C, Villar2020,Yin2020_N,Gundogan2021_NPJQI, belenchia2021quantum,gundogan2021topical}. Satellite-based quantum key distribution (SatQKD) is a precursor to long-range applications of general quantum communication~\cite{sidhu2021advances, belenchia2021quantum}. Although a general-purpose quantum network requires substantial advancements in quantum memories, multi-partite entangled state generation, routing techniques, and error correction~\cite{wehner2018quantum},  the development of SatQKD provides crucial knowledge and experience for global-scale quantum networks by developing the infrastructure and maturity of space-based long-distance quantum links.

Pioneering quantum communication demonstrations by the ${\sim}650$~kg Micius satellite showed that SatQKD and entanglement distribution is possible over record scales~\cite{jianwei2018progress, Yin2017_S, Lu2022_RMP}. Building upon these results, small satellite (${<}100$~kg) missions are attractive due to lower development costs and faster development times compared with conventional large satellites. However, the limited size, weight, and power (SWaP) available on small satellites and reduced capabilities put them at a marked disadvantage versus larger satellites such as Micius. Despite this, feasibility studies for small-satellite-based QKD and in-orbit demonstration CubeSat-based pathfinder missions are promising~\cite{Villar2020, Islam2022finite}. For low-Earth orbit (LEO) satellites, a particular challenge is the limited time window to operate a quantum channel with an optical ground station (OGS)~\cite{sidhu2021key, Sidhu2023satellite}. This limitation disproportionately constrains the volume of secure keys that can be generated due to a pronounced impact of statistical uncertainties in estimated parameters. Together with the constrained SWaP available, small-satellite missions operate under the framework of finite-resource quantum information. Understanding the impact of these constraints on SatQKD has received little attention and has both immediate and practical relevance to future satellite-based missions. Here, we fill this gap by establishing practical performance bounds on SatQKD operation under a representative set of physical resources. 

The first constraint we consider is the limited practicality of reconfiguring all QKD protocol parameters in-flight and on a pass-by-pass basis. SatQKD modelling often does not consider this, optimising the secret key length (SKL) over the entire parameter space of the protocol for each pass scenario~\cite{Sidhu2022_npjQI, Sidhu2021arxiv}. It is more realistic to consider a number of parameters as fixed, that include the operating basis bias at the OGS and the transmitted intensities. Parameter fixing has been explored in the context of terrestrial free-space QKD~\cite{airqkd2022}. In SatQKD the highly variable channel losses in SatQKD with fixed parameters require more sophisticated modeling and analysis. The limited transmission times of SatQKD further make these effects more pronounced, highlighting the importance of considering limited system adaptability. We consider a second constraint from small satellite SWaP envelopes that may limit the quantum random number generation (QRNG) subsystem driving a prepare and measure source. This directly impacts the achievable SKL by limiting signal transmission.

We start with an overview of our SatQKD system modelling and the protocol optimisation in section~\ref{sec:background}. Given the recent progress of SatQKD sources, we explore the effect of the repetition rate on key length in section~\ref{sec:source_rate}. Here, we highlight the impact of finite-key effects and establish minimum source rates based on tolerance to operational losses. Given the difficulty of implementing a SatQKD system where all parameters can be reconfigured for different overpasses, section~\ref{sec:param_fixing} explores the impact of fixed parameters on the key length. In particular, we fix the signal intensities and the receiver basis bias. In section~\ref{sec:mem_buffer}, we explore SKL generation for restricted QRNG resources and illustrate the significant impact of limited random bit generation rates on the SKL. We also determine the minimum memory storage required for non-zero finite key extraction for one overpass. Section~\ref{sec:int_fluc} explores the impact of intensity uncertainties due to limited onboard monitoring accuracy. Conclusions and discussions are provided in section~\ref{sec:conc}, where we provide key conclusions to help overcome these limitations for future SatQKD systems.


\section{Background and system model}
\label{sec:background}

\noindent
In this section, we detail our method to model channel losses, how to determine the SKL, and the optimisations considered in this work. The secret key length (SKL) achieved with the efficient BB84 protocol from a single overpass is calculated taking into account finite block size effects. 


\subsection{System model}
\label{sec:system_model}

\begin{figure}
    \centering
    \includegraphics[width=\columnwidth]{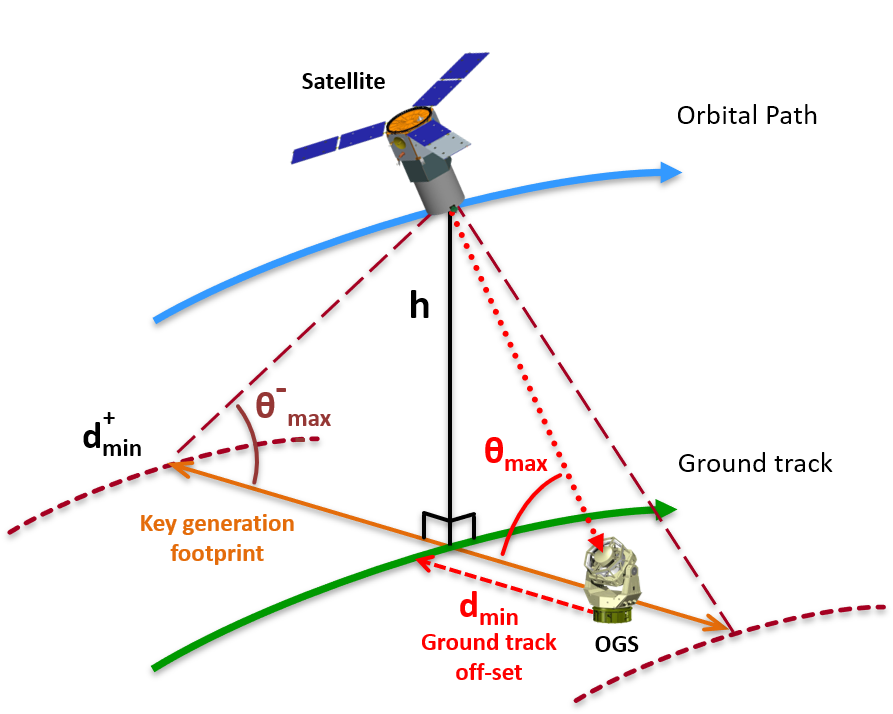}
    \caption{\textbf{General satellite overpass geometry.} The satellite reaches a maximum elevation of $\theta_\text{max}$, corresponding to the minimum OGS ground track distance, $d_\text{min}$. The smallest $\theta_\text{max}$ that generates a non-zero finite key is denoted $\theta_\text{max}^-$ and characterises the operational SatQKD key generation footprint $2d^+_\text{min}$.}
    \label{fig:geom}
\end{figure}%

\noindent
We consider a satellite in a circular Sun-synchronous orbit (SSO) of altitude $h=500$~km implementing downlink QKD to an OGS during the night to minimise background light. The elevation and range of the satellite-OGS channel are calculated as a function of time for different satellite overpass geometries and ground track offsets, $d_\text{min}$, and maximum satellite overpass elevations, $\theta_\text{max}$ (Fig.~\ref{fig:geom}). Different satellite overpasses have different values for $d_\text{min}$. This means $d_\text{min}$ can be used to characterise each overpass. In fact, for a fixed orbital altitude, the ground track offset $d_\text{min}$ and the maximum elevation angle, $\theta_\text{max}$, are equivalent. The ideal overpass corresponds to the satellite passing the OGS directly overhead, or zenith ($d_\text{min} = 0$ m, $\theta_\text{max}=90^\circ$), since it provides the longest transmission time and has the lowest average channel loss. Generally, a satellite will not pass zenith but will reach a maximum elevation $\theta_\text{max}({<}90^\circ)$. We consider a minimum elevation transmission limit of $\theta_\text{min}=10^\circ$ that reflects practicalities such as local horizon visibility and system pointing limitations.

The instantaneous link efficiency depends on the elevation $\theta(t)$, the range $R(t)$ between the satellite and OGS, and source wavelength $\lambda$, and is used to generate count statistics. For a fixed orbital altitude, the satellite-OGS range is implicitly defined through the satellite's elevation. The link efficiency is then defined as (in dB),
\begin{align}
  \eta_{\lambda}\left(\theta\right) = \eta_\text{diff}\left(\lambda,\theta\right) + \eta_\text{atm}\left(\lambda,\theta\right)
    + \eta_\text{int},
\label{eq:ins_loss_func}    
\end{align}
where $\eta_\text{diff}$, $\eta_\text{atm}$, and $\eta_\text{int}$ are losses from diffraction, atmospheric scattering and absorption, and a fixed `intrinsic' system efficiency respectively. To characterise the overall system electro-optical efficiency independent of satellite overpass trajectory, we define the system loss metric, $\slm$, as the total instantaneous link efficiency at zenith. Diffraction losses are estimated using the Fraunhofer approximation to the Rayleigh-Sommerfeld diffraction integral to determine the power at the receiver, $P_R$, 
which is normalised by the power at the transmitter, $P_T$ such that $\smash{\eta_\text{diff} = -10 \log_{10}(P_R/P_T)}$. Atmospheric absorption and scattering losses are calculated using $\smash{\eta_\text{atm} = -10 \log_{10}T_{\lambda}}$, where the transmissivity, $T_{\lambda}$, is determined using MODTRAN for a given wavelength and elevation~\cite{Modtran_inproceedings}. The `intrinsic' system loss, $\eta_\text{int}$, accounts for: fixed losses inherently built into the system due to detector efficiency, internal losses of the receiver; pointing losses; and imperfect non-diffraction-limited beam propagation, and is conservatively set to 20~dB to model a SatQKD system with overall $\slm = 40$~dB. Different SatQKD systems with various fixed losses can be modelled by scaling the $\slm$ value. See Methods~\ref{subsection:loss_modelling} for more detail on loss modelling.

\begin{figure}
    \centering
    \includegraphics[width=\columnwidth]{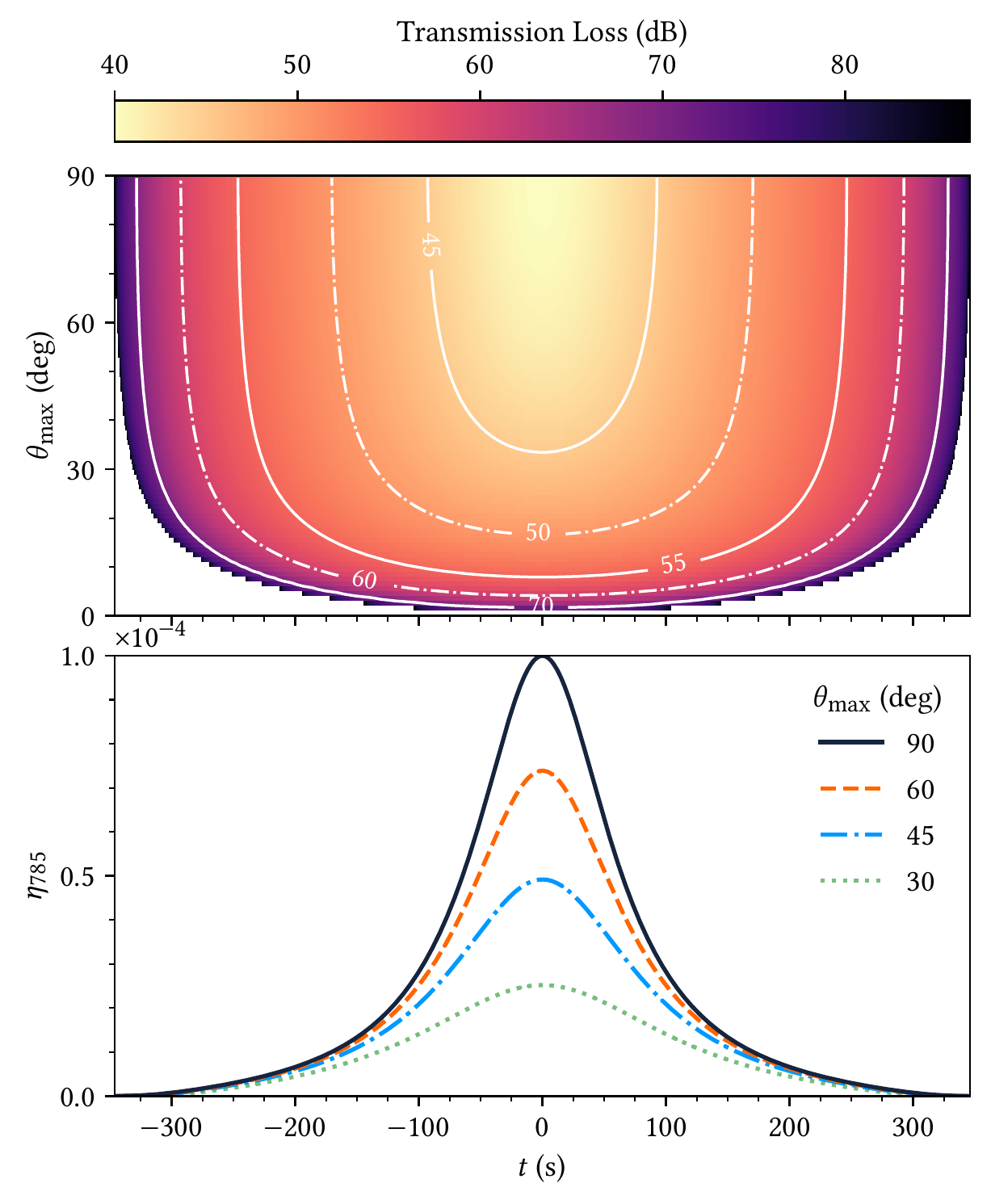}
    \caption{\textbf{Link model for satellite-to-ground QKD}. (top) Instantaneous link efficiency, $\eta_{785}$ (Eq.~\eqref{eq:ins_loss_func}), for different satellite overpasses with maximum elevation, $\theta_\text{max}$, and time with $\lambda=785$ nm. The smallest transmission loss of 40 dB occurs for a zenith overpass ($\theta_\text{max} = 90^\circ$) at time $t=0$. (bottom) $\eta_{785}$ for specific $\theta_\text{max}$. System parameters as in Table~\ref{tab:system_parameters}.}
    \label{fig:loss}
\end{figure}%

The link loss characterises the probability that a single photon transmitted by the satellite is detected by the OGS. A lower dB value of $\eta_\text{link}$ represents smaller loss due to better system electro-optical efficiency. This improvement could stem from the use of larger transmit and receive aperture diameters, better pointing accuracy, lower receiver internal losses, and higher detector efficiencies. Internal transmitter losses are not included since they can be countered by adjusting the weak coherent pulse (WCP) source to maintain the desired exit aperture intensities~\cite{Bourgoin:2013fk}. We also do not explicitly consider time-varying transmittance, modelling the average change in channel loss due only to the change in elevation with time. For discrete variable QKD (DV-QKD) protocols, e.g. BB84, channel transmissivity fluctuations do not directly impact the secret key rate, in contrast to continuous variable QKD where this appears as excess noise leading to key reduction~\cite{Usenko2012_NJP,hosseinidehaj2020composable}.

We model a small satellite QKD system, for example~\cite{colquhoun2022responsive}, implementing a decoy-state BB84 protocol in a downlink configuration for QKD service provision using a WCP source. We consider a source wavelength of $\lambda=785$~nm, a transmitter (receiver) aperture diameter of 8~cm (70~cm), and a Gaussian beam waist of 8~cm. Our general analysis is wavelength agnostic, but we specifically analyse $\lambda=785$~nm as this is representative of several missions currently in development~\cite{Podmore2021qkd,Islam2022finite, colquhoun2022responsive}, partly due to favorable atmospheric transmission and the availability of relevant sources and detectors~\cite{Bourgoin:2013fk}. Fig.~\ref{fig:loss} illustrates the modelled transmission loss and link efficiency for different overpass geometries. 

In addition to this link loss, we include several error sources. First, after-pulsing in a photon detector can have adverse effects on the estimate of click statistics. While the after-pulsing probability is detector and operating condition dependent, we take a value of 0.1\%, which is consistent with the literature~\cite{Hwang2003_PRL,Chen2021,Zhang2017_PRA}. Second, the intrinsic quantum bit error rate, $\text{QBER}_\text{I}$, is defined as the lumped error from source quality, receiver measurement fidelity, basis misalignment, and polarisation fluctuations~\cite{Toyoshima2009_OE}. Finally, we define the extraneous count probability, $p_\text{ec}$, as the sum of dark and background light count rates and is assumed constant and independent of elevation. Together, these losses and errors provide a complete characterisation of a SatQKD system and are summarised in Table~\ref{tab:system_parameters}.

Before concluding this section, we note that our current analysis could be extended to model an uplink channel by using a suitable link-loss model (loss vs elevation). A ground-to-satellite link will increase channel losses due to the shower curtain effect. While turbulence is highly dependent on elevation, it generally leads to an additional 20~dB of loss compared to a downlink channel~\cite{Bourgoin:2013fk}.

\newcommand*{\tabindent}{ \hspace{-1mm}}
\newlength{\thickarrayrulewidth}
\setlength{\thickarrayrulewidth}{2.1\arrayrulewidth}
\renewcommand{\arraystretch}{1.25}
\setlength{\tabcolsep}{8pt}
\begin{table}[t!]
  \centering
  \begin{tabular}{m{4cm}|m{1.4cm}|m{1.4cm}}
    \thickhline
    \textbf{Parameter description} 				& \textbf{Notation}				& \textbf{Value} 	\\
    \hline
    Transmitter aperture diameter  				&	$T_X$					& 8 cm			\\
    Receiver aperture diameter					&	$R_X$					& 70 cm			\\
    Gaussian Beam waist						&	$w_0$					& 4 cm			\\
    Source wavelength						&	$\lambda$					& 785 nm			\\    
    Source rate		 						&	$f_s$					& 500 MHz		\\ 
    Satellite orbit altitude	 					&	$h$						& $500$ km 		\\ 
    Minimum elevation limit						&	$\theta_\text{min}$			& $10^\circ$		\\
    Intrinsic quantum bit error rate				& 	$\text{QBER}_\text{I}$		& 0.5\% 			\\ 
    Extraneous count probability					&	$p_\text{ec}$				& $5\times 10^{-7}$ 	\\ 
    After-pulsing probability	  					&	$p_\text{ap}$				& 0.1\% 			\\ 
    System loss metric						&	$\eta^\text{sys}_\text{loss}$	& $40$ dB 		\\ 
    \textcolor{black!60!white}{
    $\hookrightarrow$Diffraction loss at zenith}		&	\textcolor{black!60!white}{$\eta_\text{diff}(\lambda, 90)$}	& \textcolor{black!60!white}{19.4~dB}		\\
    \textcolor{black!60!white}{$\hookrightarrow$Atmospheric loss at zenith}	&	\textcolor{black!55!white}{$\eta_\text{atm}(\lambda, 90)$}	& \textcolor{black!60!white}{0.6~dB}		\\
    \textcolor{black!60!white}{$\hookrightarrow$Optical inefficiency}		&	\multirow{2}{*}{\textcolor{black!60!white}{$\eta_\text{int}$}}				& \textcolor{black!60!white}{$12.0$~dB} 		\\ 
    \textcolor{black!60!white}{$\hookrightarrow$Imperfect beam propagation}	&					& \textcolor{black!60!white}{$8.0$~dB} 		\\ 
    \color{black}
    Correctness parameter 						&	$\epsilon_c$				& $10^{-15}$ 		\\ 
    Security parameter						&	$\epsilon_s$				& $10^{-10}$ 		\\     
    \thickhline
  \end{tabular}
  \caption{\textbf{Reference system parameters}. Transmitter, receiver, and source properties determine range and elevation-dependent loss. The system loss metric, $\eta^\text{sys}_\text{loss}$, defined as the link efficiency at zenith, is 40~dB. The `intrinsic' system loss is broken down into two components (Methods~\ref{subsection:loss_modelling}). $\eta^\text{sys}_\text{loss}$ can be scaled to model other SatQKD systems that differ by a fixed link loss ratio, e.g. different $T_X$ or $R_X$ apertures, or detector efficiencies. The intrinsic quantum bit error rate, $\text{QBER}_\text{I}$, incorporates errors from source quality, receiver measurement fidelity, basis misalignment, and polarisation fluctuations, while the extraneous count probability, $p_\text{ec}$, incorporates detector dark count and background rate. The correctness and security parameters are used to determine the finite-block composable SKL.}
  \label{tab:system_parameters}
\end{table}%
%


\subsection{The protocol and secret key length}
\label{subsec:protocol}

\noindent
The QKD protocol we investigate is efficient Bennett-Brassard (BB84) with two decoy states, i.e. three different pulse intensities~\cite{Lim2014_PRA,lo2005efficient,Hwang2003_PRL,Wang2005_PRL,Lo2005_PRL,Yin2020_N}. In this protocol, the transmitter (Alice) and the receiver (Bob) encode bits within one of two polarisation bases, denoted $\mathsf{X}$ and $\mathsf{Z}$. We adopt the convention that the $\mathsf{X}$ basis is used for key bits, while the $\mathsf{Z}$-basis is used to detect an eavesdropper through the phase error rate. Alice prepares bits in the $\mathsf{X}$-basis with probability $P^A_\mathsf{X}$, while Bob measures within the $\mathsf{X}$-basis with probability $P^B_\mathsf{X}$. It is standard to take $P^A_\mathsf{X}=P^B_\mathsf{X}=P_\mathsf{X}$, however, in general it is possible that $P^A_\mathsf{X} \ne P^B_\mathsf{X}$, particularly if one probability is fixed due to practical considerations~\cite{airqkd2022}. We consider phase-randomised coherent pulses where the intensity (mean photon number) $\mu_k\in\{\mu_1,\mu_2,\mu_3\}$ is randomly chosen with probability $p_{\mu_k}$. There are alternative carriers to phase-randomised coherent pulses. True single-photon sources could be considered~\cite{Morrison2022_arxiv, Juboori2023_arxiv, MurtazaOE_2023, Abasifard2023_arxiv}, amongst others~\cite{QKDreview2020}, though these are at a much lower stage of maturity, for terrestrial or space applications, compared with WCP sources.

After the quantum signals are transmitted from Alice to Bob, they perform a standard reconciliation procedure to correlate detection events with transmitted pulses, basis matching, intensity announcement, and parameter estimation. Only the bits in the $\mathsf{X}$-basis are used for the key, while the $\mathsf{Z}$-basis bits are made public. The raw key is formed by performing error correction on the $\mathsf{X}$-basis bits, which necessitates the public exchange of $\lambda_\text{EC}$ bits in the information reconciliation phase. In practice, the value of $\lambda_\text{EC}$ is known from the error correction communication, but for the purposes of modelling we use an estimate that varies with the block size, quantum bit error rate, and the required correctness parameter~\cite{Tomamichel2017_QIP}. This estimate generates suitable values for the error correction efficiency for SatQKD data representative of current engineering efforts and capabilities (see Methods~\ref{subsec:error_corr_term} for a detailed discussion and demonstration). The results for the $\mathsf{Z}$-basis are used to estimate parameters such as the number of bits from vacuum events, $s_{\mathsf{X},0}$, the number of bits from single photon events $s_{\mathsf{X},1}$, and the phase error $\phi_{\mathsf{X}}$. The exact formulas for these terms are provided in Ref.~\cite{Sidhu2022_npjQI}, which is based on Refs.~\cite{Lim2014_PRA,Yin2020_N}. After privacy amplification, the final SKL, $\ell$, is given by~\cite{Lim2014_PRA}
\begin{align}
    \ell  = \Big\lfloor s_{\mathsf{X},0}  + s_{\mathsf{X},1} (1 - h(\phi_\mathsf{X}))
     - \lambda_{\text{EC}} - 6 \log_2 \frac{21}{\epsilon_\text{s}} - \log_2 \frac{2}{\epsilon_\text{c}}\Big\rfloor,
\label{eqn:skl_lim_result}    
\end{align}
where $h(x)=-x\log_2(x)-(1-x)\log_2(1-x)$ is the binary entropy function, and $\epsilon_{\text{s}}$ and $\epsilon_{\text{c}}$ are the composable security and correctness parameters respectively~\cite{Lim2014_PRA,Renner2006_thesis}.

We can maximise the SKL, Eq.~\eqref{eqn:skl_lim_result} by optimising over the protocol parameters $p_k$, $\mu_k$, and $P_\mathsf{X}$ for a given satellite-OGS overpass, system link efficiency, and system configuration (as in Table~\ref{tab:system_parameters}). The value of $\mu_3$ is set to vacuum since this helps with the estimate of the vacuum counts, $s_{\mathsf{X},0}$~\cite{Lim2014_PRA}. The transmission time window from which the finite block is constructed is an additional important optimisation parameter to maximise the achievable finite key~\cite{Sidhu2022_npjQI}. This is because, under finite-size security analysis, higher QBER increases the minimum raw key length necessary for non-zero key length extraction due to less efficient reconciliation and post-processing overheads. However, taking the largest block size permitted by a satellite overpass is sometimes not the best strategy. This is since data from lower elevations have both smaller count rates and higher signal QBER, which increases the average channel QBER and may offset any improvements to the SKL from larger block sizes. We define the processing block transmission time window to run from $-\Delta t$ to $+\Delta t$, such that the total transmission time is $2\Delta t$ with $t=0$ corresponding to the time of highest elevation $\theta_\text{max}$. The SKL in Eq.~\eqref{eqn:skl_lim_result} is additionally optimised over discretised values for $\Delta t$, and the value for $\Delta t$ chosen that yields the largest SKL. This full optimisation is performed in version 1.1 of the Satellite Quantum Modelling and Analysis (SatQuMA) software~\cite{Sidhu2021arxiv}. For more details on the software and the numerical optimisation see Refs.~\cite{Sidhu2022_npjQI, Sidhu2021arxiv}.

This fully optimised scenario yields an upper bound to SatQKD performance. In practice, these bounds may be difficult to achieve due to constraints and trade-offs in the mission design and operation. In the following section, we provide an overview of modifications to the optimisation problem with constraints that closely reflect operational considerations for the derivation of realistic performance bounds.


\subsection{Practical optimisation of the secret key length}
\label{subsec:optimisation}

\noindent
The original protocol parameter optimisation problem is modified to handle different numerical investigations. Though classical communication constraints are important for SatQKD operations, we do not consider these limitations (see Ref.~\cite{Sidhu2022_npjQI} for a brief discussion). First, section~\ref{sec:source_rate} introduces the source-rate normalised SKL to illustrate the impact of finite-key effects on the SKL and to provide an informed decision on the source rate to consider for the remainder of the work. Second, section~\ref{sec:param_fixing} fixes the values of the signal intensity $\mu_1$, decoy intensity $\mu_2$, and the receiver basis bias $P^B_\mathsf{X}$, since it may not be practical to change these parameters on a pass-by-pass basis in an operational system. The transmitter and receiver basis biases are allowed to differ, i.e. $P^A_\mathsf{X} \ne P^B_\mathsf{X}$, to model a fixed OGS basis bias and adjustable transmitter bias. The SKL is then maximised over the remaining protocol parameter space defined by the set $\{P^A_\mathsf{X}, p_{\mu_1}, p_{\mu_2}, \Delta t\}$. The fixed values for $P^B_\mathsf{X}$, $\mu_1$, and $\mu_2$ are set to those that maximise the expected annual SKL through a procedure detailed in Methods~\ref{subsection:param_fixing}. Third, section~\ref{sec:mem_buffer} explores the impact of QRNG subsystem limitations that may constrain the number of signals that can be transmitted during an overpass. This is modelled using a finite-sized onboard random number memory store, corresponding to an associated transmission cutoff time, from which we determine the reduction in long-term average key generation rate. We also determine the minimum memory buffer required to generate non-zero SKL. Finally, in section~\ref{sec:int_fluc}, we consider the effect of pulse intensity uncertainties on the secure key that can be extracted taking into account reduced intensity knowledge. For this, the signal and decoy state intensities are sampled between a range that depends on the uncertainty percentage of the intended intensity values.


\section{Results}
\subsection{Source rate}
\label{sec:source_rate}

\noindent
Micius performed finite key generation with a 100 MHz source repetition rate, later upgraded in-flight to 200~MHz~\cite{Chen2021}. Miniaturisation of such high-speed sources enables their use on small satellites. For example, increasing the source repetition rate leads to a larger block size that reduces statistical uncertainties in parameter estimation, hence a higher finite key rate. This expands the pass opportunities that result in non-zero secret keys, enhancing the robustness and effective key transmission footprint of a SatQKD system~\cite{Sidhu2022_npjQI}. In addition, the use of high-speed sources can help higher altitude SatQKD operation by partially compensating for increased channel losses~\cite{Sidhu2022_npjQI}. In this section, we investigate the effect of operating source rate, $f_s$, on the robustness of SatQKD systems to channel loss in the finite key regime. 

\begin{figure}
    \centering
    \includegraphics[width=\columnwidth]{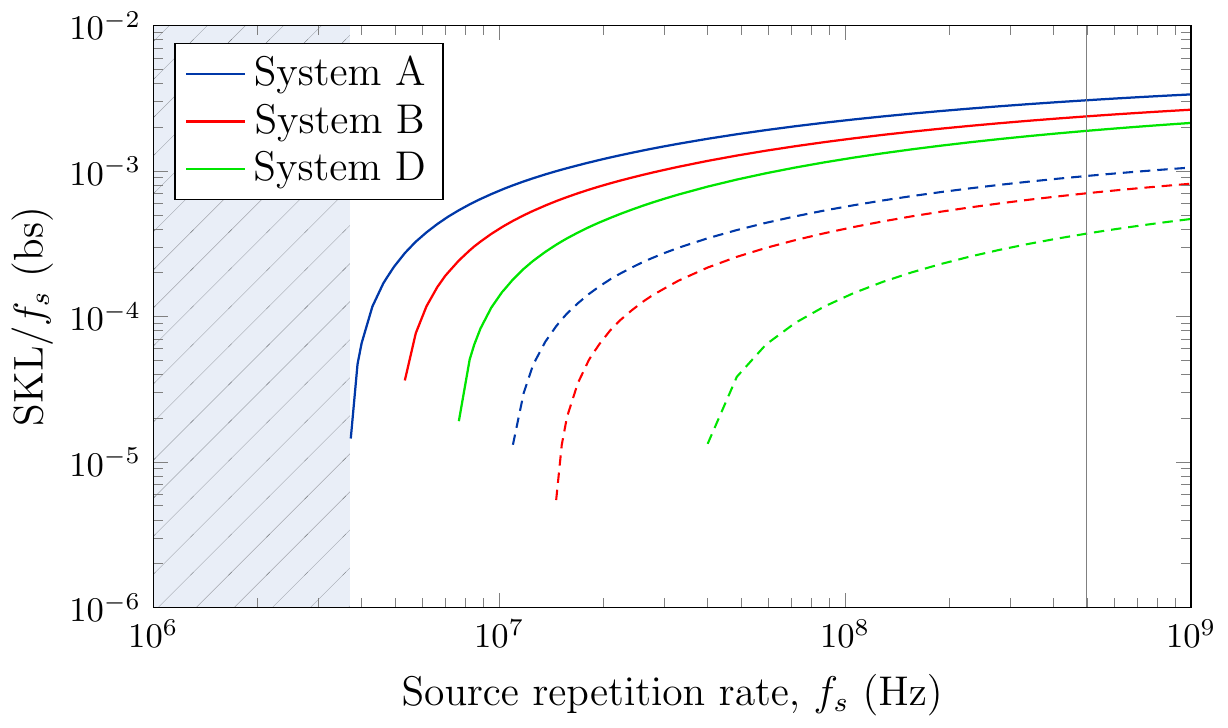}
    \caption{\textbf{Finite key efficiency vs source rate}. Source rate normalised SKL as a function of $f_s$ for overpasses with $\theta_\text{max} = 90^\circ$ (solid lines) and $30^\circ$ (dashed lines), for three system configurations $\{\text{QBER}_\text{I}, p_\text{ec}\}$: A = $\{0.1\%, 1\times10^{-8}\}$, B = $\{0.5\%, 1\times10^{-8}\}$, and D = $\{0.5\%, 1\times10^{-7}\}$. The critical $f_\text{s}$ value corresponds to the transition of zero and non-zero finite SKL. The shaded blue region illustrates the key suppression region for System A with $\theta_\text{max}=90^\circ$ where statistical fluctuations in estimated parameters overwhelm key generation due to finite available statistics. The vertical line is at $f_s = 500$~MHz, which we consider for the remainder of the paper.}
    \label{fig:source_vary}
\end{figure}

To evaluate finite key efficiency, Fig.~\ref{fig:source_vary} illustrates the source rate normalised SKL as a function of source rate for a zenith overpass (solid lines) and a satellite overpass with $\theta_\text{max}=30^\circ$ (dashed lines) for three different system configurations of $\{\text{QBER}_\text{I}, p_\text{ec}\}$. For a given time window $\Delta t$, the block size increases with increasing $f_s$, which improves the normalised finite SKL. This improvement indicates a critical value $f_s^\text{crit}$ below which finite key effects overwhelm raw key transmission and the distillable finite SKL is zero. For $f_s < f_s^\text{crit}$, this \emph{key suppression region} is illustrated in shaded blue for System A with $\text{QBER}_\text{I}=0.1\%$, $p_\text{ec}=1\times10^{-8}$, and $\theta_\text{max}=90^\circ$. Above $f_s^\text{crit}$, we note the SKL scales super-linearly with the source rate due to  multiple improvements in parameter estimation, error correction efficiency, and reduced overhead of the composable security parameters with increasing block length.

The vertical gray line in Fig.~\ref{fig:source_vary} corresponds to $500$ MHz, well outside the key suppression region, that we take as a representative value for a near-term small satellite source. This provides robustness against a range of typical extraneous counts and intrinsic QBERs expected in SatQKD and provides feasible finite key generation for a single satellite overpass, but is compatible with modest receiver detectors. Higher source rates, though providing larger key lengths, require lower detector timing jitter. Silicon single-photon avalanche photodiodes (Si-SPADs) typically have timing jitter in the order of $\sim$ 0.5~ns~\cite{Ceccarelli2021_AQT} compatible with coincidence windows of $\sim$ 1~ns and interpulse separations of 2~ns. Extending clock rates to the GHz range requires lower timing jitters such as provided by superconducting nanowire single-photon detectors (SNSPDs)~\cite{Holzman2019_AQT} at the expense of greater SWaP and cost (SWaP-C) owing to the need for cryogenic operation and single mode coupling that raises further system design issues. Therefore, the following analysis will assume a source rate of $500$~MHz unless stated otherwise given it balances the tradeoff between detector performance requirements, hence SWaP-C, and count rate.


\subsection{Impact of parameter fixing}
\label{sec:param_fixing}

\begin{figure*}[t!]
    \centering
    \includegraphics[width=0.95\linewidth]{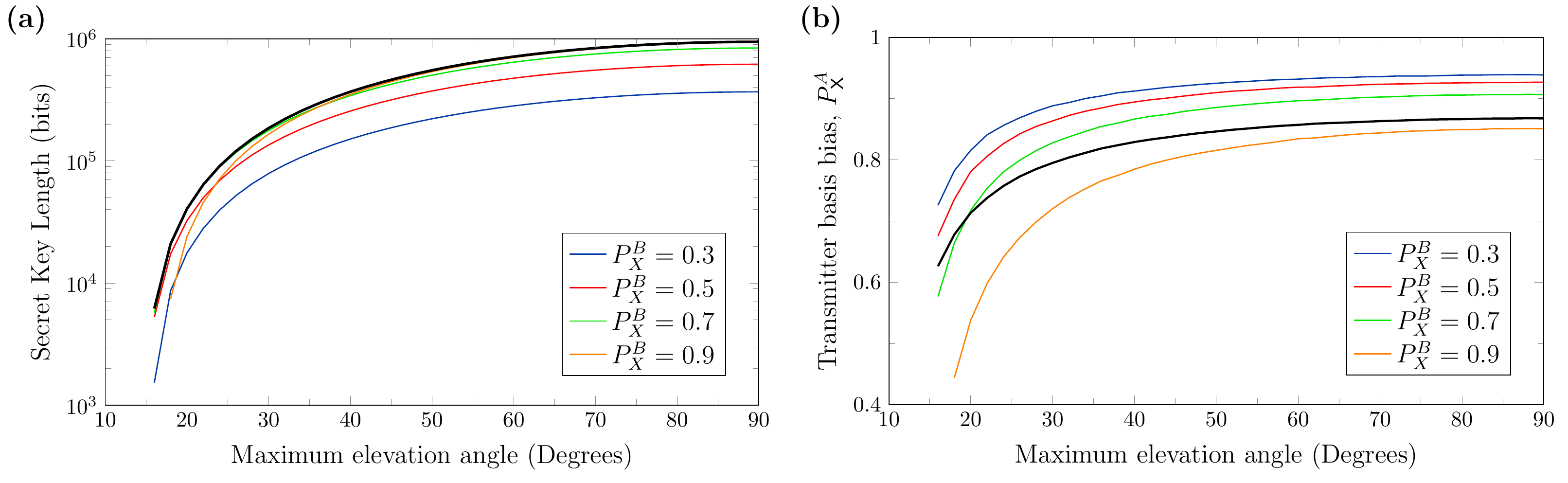}
    \caption{\textbf{Impact of fixed receiver basis bias and source intensities}. All curves are for $\mu_1=0.71$, $\mu_2=0.14$, $\mu_3=0$, $p_\text{ec}=10^{-7}$ and $\text{QBER}_\text{I}=0.005$. (a) SKL as a function of $\theta_\text{max}$ for a fixed $\smash{P^B_\mathsf{X}}$ and fixed pulse intensities. For reference, the black solid line represents the optimal SKL maximised over $\smash{P^A_\mathsf{X}}$ and $\smash{P^B_\mathsf{X}}$ with the same fixed intensity values. (b) Plots of optimised values for $\smash{P^A_\mathsf{X}}$ as a function of $\theta_\text{max}$ for a fixed basis $\smash{P^B_\mathsf{X}}$ and fixed pulse intensities. The black solid line represents the optimal basis bias $\smash{P_\mathsf{X}^A}$ with the same fixed intensity values.} 
    \label{fig:param_fixing_plots}
\end{figure*}%

\noindent
SatQKD modelling often involves optimising the operational parameter space associated with the protocol and system configuration to maximise the number of finite keys generated. However, achieving these optimised key lengths assumes all parameters can be easily changed to operate at their optimised values. It may be desirable on cost, complexity, and robustness grounds to deploy SatQKD systems with limited reconfigurability, motivating analyses where some parameters are fixed. First, the OGS basis choice is often implemented passively using a fixed beamsplitter. Thus, changing receiver basis bias by physically swapping out the beamsplitter for different optimised values on a per-pass basis may be impractical in live deployment. A variable beamsplitter could be considered but with cost, complexity, and performance considerations. Note that the transmitter basis bias can be easily adjusted in the random bit generation and processing of the data used to control the source, hence we consider this parameter to be easily varied. Second, all the operational pulse intensities $\mu_j$ may be fixed pre-flight to avoid more complex source driving systems with increased SWaP-C and reliability concerns. Since the optimal decoy-state intensities strongly depend on the channel loss, background counts, and the satellite's orbital trajectory, fixed values may significantly impact the SKL.

In this section, we determine the impact of these engineering constraints on the finite SKL. We constrain the receiver basis bias and decoy-state intensities to certain fixed values, such that $P^B_\mathsf{X}=\{0.3, 0.5, 0.7, 0.9\}$ (commonly available beamsplitter splitting ratios) in addition to the ideal value of $P_\mathsf{X}^B = 0.84$ that corresponds to a custom beamsplitter and $\{\mu_1, \mu_2, \mu_3\}=\{0.71, 0.14, 0\}$. The derivation of these ideal values can be seen in Methods~\ref{subsection:param_fixing} for fixed parameter optimisation that maximise the long-term average SKL. For these fixed values, Fig.~\ref{fig:param_fixing_plots}(a) illustrates the finite SKL as a function of different satellite overpasses. Despite this restriction, we note it is possible to generate near-optimal SKLs across a wide range of elevation angles. Further, increasing the OGS bias can generate higher finite SKL. However, we observe that for a choice of $P^B_\mathsf{X}=0.9$, it is not possible to extract a secret key at lower $\theta_\text{max}$. This suggests that choosing too large an OGS bias can reduce the key generation capacity, owing to fewer overpasses opportunities that generate a non-zero key. To understand this effect, we recall that a larger receiver basis bias corresponds to a smaller portion of received bits dedicated to parameter estimation. Therefore, choosing a large OGS basis bias at larger average channel QBERs leads to less efficient parameter estimation, which generates zero secret keys. SatQKD systems should therefore carefully choose the fixed OGS bias to address the tradeoff between a maximised single pass SKL and the long-term key generation capacity. Notice that the secret key length for $P^B_\mathsf{X}=0.7$ is approximately the same as for $P^B_\mathsf{X}=0.9$, but with non-zero keys at lower elevations.

Fig.~\ref{fig:param_fixing_plots}(b) illustrates the optimal $P^A_\mathsf{X}$ values that maximise the SKL as a function of elevation angle for each fixed value of the receiver basis bias. We first note the basis bias for the transmitter and receiver are generally different, which differs from the usual case considered in the literature. The value of $P^A_\mathsf{X}$ can vary to compensate for the fixed value of $P^B_\mathsf{X}$. One can show that if both $P^B_\mathsf{X}$ and $P^A_\mathsf{X}$ can vary freely, then the optimal raw key length is found for $P^B_\mathsf{X}=P^A_\mathsf{X}$~\cite{airqkd2022}. From Fig.~\ref{fig:param_fixing_plots}(b) we find that for $P^B_\mathsf{X}=0.3$ and 0.5, we observe that $P^A_\mathsf{X} > P^B_\mathsf{X}$. This suggests that a small fixed receiver basis bias leads to too large a portion of signals dedicated to parameter estimation, which is compensated for by choosing a large transmitter basis bias. Equally, for $P^B_\mathsf{X}=0.9$ we observe that $P^A_\mathsf{X} < P^B_\mathsf{X}$. This clearly demonstrates that when we \emph{fix} $P^B_\mathsf{X}$, then choosing an equal basis bias is not optimal. However, when we are free to optimise \emph{both} $P^B_\mathsf{X}$ and $P^A_\mathsf{X}$, then choosing $P^A_\mathsf{X}=P^B_\mathsf{X}$ is optimal~\cite{airqkd2022}.

Despite the impracticality of implementing a fully optimised parameter space, we find a number of ways SatQKD missions can enhance finite key generation. This involves careful selection of $P^B_\mathsf{X}$ that maximises both the single-pass SKL and the long-term key generation capacity and careful selection of the decoy-state intensities that can counter the effects of large channel losses.


\subsection{QRNG subsystem limitations}
\label{sec:mem_buffer}

\noindent
Prepare and measure protocols require random bits for the preparation of signal states. QRNGs with the required rate to feed a high-speed source in real time may incur significant onboard processing resources and SWaP. Alternatively, the random bits can be generated at a much slower rate with less resource-hungry QRNGs prior to the overpass, assuming that the transmission time duty cycle is small compared to the total orbital time. For this latter situation, we consider limits on the amount of onboard storage for random bits to drive the source, often limited on small satellites. This constrains the amount of reconciled data established between a satellite and OGS, thus directly impacting the achievable SKL per pass. Unlike in previous sections where we assumed the source can run indefinitely, in this section, we extend our analysis to model the impact of varying memory storage limits of cryptographically secure random bits on the final SKL.

For a two decoy-state weak coherent state protocol, each pulse consumes four random bits; one for the basis choice, one for the key value, and two for the intensity choice. For the efficient BB84 decoy-state protocol, the basis choice bit and the intensity bits are biased. In general, it takes at most two unbiased bits on average to generate one biased bit~\cite{gryszka2021biased}, hence each pulse requires up to seven unbiased bits from the quantum random number generator (QRNG), though only four bits need to be stored after biasing. At 500~MHz source rate, this requires 2~Gb/s of stored random bits to drive the source. Therefore, a zenith pass with a maximum overpass duration of 444~s (accounting for a minimum elevation limit of 10$^\circ$) requires a minimum availability of 111~GB of random bits. Current state of the art in space-validated QRNGs  can achieve rates of 1-20~Mb/s~\cite{QRNG2016,IDquantQRNG}, which falls short to support complete transmission, and thus necessitates a buffer.

\begin{figure}[t!]
    \centering
    \includegraphics[width=\linewidth]{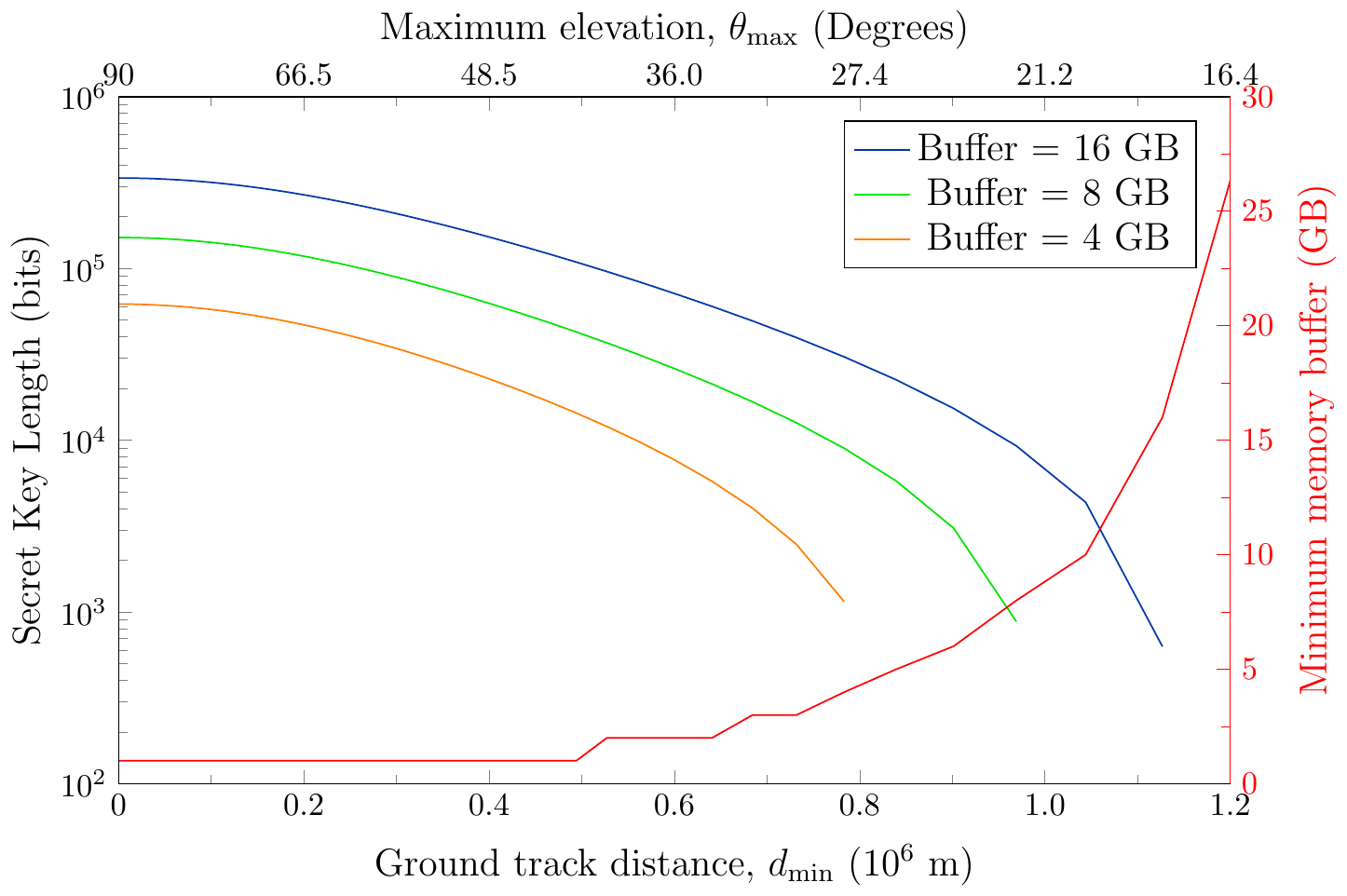}
    \caption{\textbf{Overpass and memory buffer effects.} SKL (left axis) and minimum memory buffer (right axis) as a function of ground track distance. We consider $\slm=40$~dB, $f_\text{s}=500$~MHz, $\text{QBER}_\text{I}=0.5\%$, and $p_\text{ec}=1\times10^{-7}$. A larger memory buffer permits a longer transmission time, which extends the operational footprint of the SatQKD system. Further, a larger minimum memory buffer requirement is observed at larger ground track distances to generate non-zero finite keys. This provides an indication of SatQKD system specifications.}
    \label{fig:membuff}
\end{figure}

First, we examine the effects of a limited random bit memory buffer on the finite key.  An 8~GB buffer can support up to 32~s transmission time for a 500~MHz source, which is much shorter than the maximum overpass duration of 444~s. Fig.~\ref{fig:membuff} (left-hand axis) shows the per-pass SKL for different memory buffers as a function of overpass geometry ($d_\text{min}$, $\theta_\text{max}$). A larger memory buffer permits longer transmission times, which enhances the finite SKL and extends the operational footprint of the SatQKD system. Second, we determine the minimum memory buffer required to yield non-zero finite keys for different overpasses. For a given overpass, the smallest block size that yields a non-zero finite key defines the smallest operational time window, $t_\text{min}$, that should be supported by the onboard storage. This provides a measure of the memory buffer requirement for a SatQKD mission, given by $f_s t_\text{min}/2$ Bytes. The right-hand axis of Fig.~\ref{fig:membuff} illustrates the minimum memory buffer required for different satellite overpass trajectories. The demand for larger onboard storage requirements increases with increasing ground track distances. This is because satellite overpasses with larger ground track distances require larger minimum transmitted signals to overcome the larger average channel losses and generate a non-zero finite key.

\begin{figure}
    \centering
    \includegraphics[width=\linewidth]{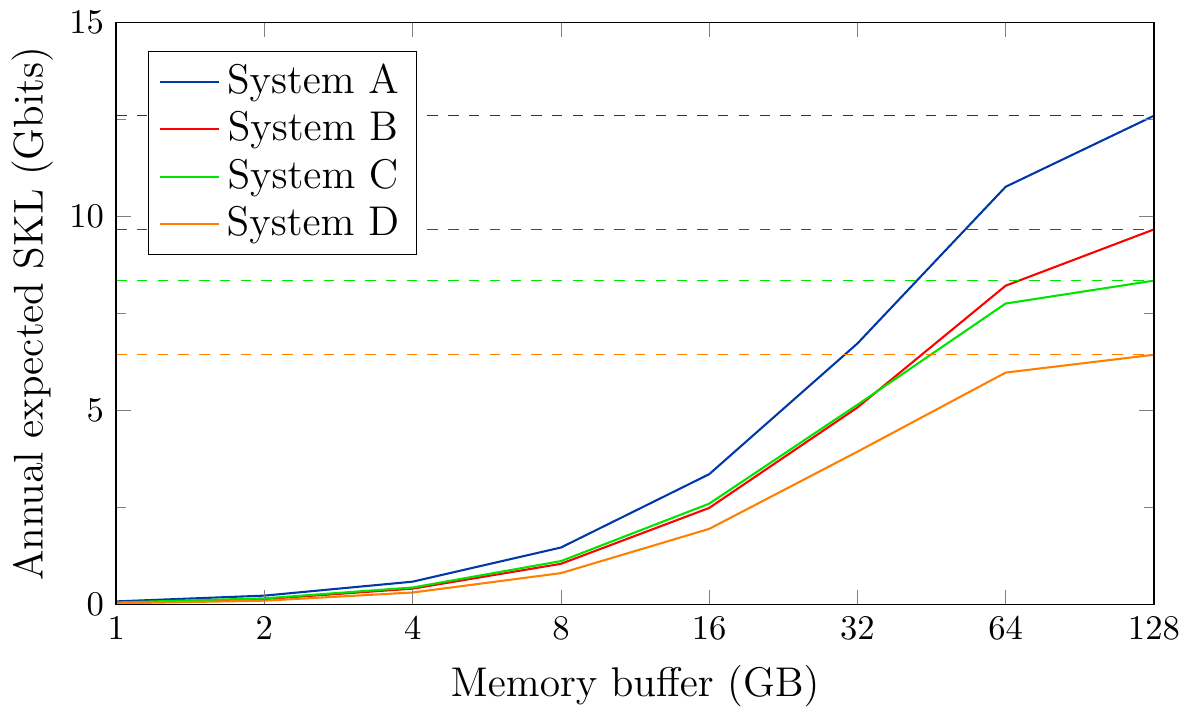}
    \caption{\textbf{Annual expected SKL vs Memory Buffer.} The OGS is at a latitude of $\bm{55.9^\circ}$~N, $f_\text{s}=500$~MHz, and $\slm=40$~dB. We illustrate four distinct system configurations $\{\text{QBER}_\text{I}, p_\text{ec}\}$: A = $\{0.1\%, 1\times10^{-8}\}$, B = $\{0.5\%, 1\times10^{-8}\}$, C = $\{0.1\%, 1\times10^{-7}\}$, and D = $\{0.5\%, 1\times10^{-7}\}$. Dashed lines indicate the annual SKL expected without memory constraints.}
    \label{fig:varymem}
\end{figure}

Third, to quantify the overall impact of limited memory buffers on the SKL, we estimate the annual amount of secret keys that can be generated using methods from Ref.~\cite{Sidhu2022_npjQI}. For a
Sun-synchronous orbit and neglecting weather effects, the expected annual key for single overpass blocks with an OGS site situated at a particular latitude is approximated by~\cite{Sidhu2022_npjQI}
\begin{align}
\overline{\text{SKL}}_\text{year}=N_\text{orbits}^\text{year}\frac{\text{SKL}_\text{int}}{L_\text{lat}},
\end{align}
where $\text{SKL}_\text{int}$ is twice the integrated area under the SKL vs $d_\text{min}$ curve in Fig.~\ref{fig:membuff} (units of bit metres), $N_\text{orbits}^\text{year}$ is the number of orbits per year, and $L_\text{lat}$ is the longitudinal circumference along the line of latitude at the OGS location. Fig.~\ref{fig:varymem} illustrates how $\overline{\text{SKL}}_\text{year}$ varies as a function of the memory buffer for an OGS at a latitude of $55.9^\circ$ N (latitude of Glasgow). For our reference configuration (System D) with $\slm=40$~dB, $\overline{\text{SKL}}_\text{year}$  is 0.81~Gb (3.94~Gb) for a memory buffer of 8~GB (32~GB) respectively. For comparison, without QRNG limitations, $\overline{\text{SKL}}_\text{year}$ is 6.44~Gb. Fig.~\ref{fig:varymem} also shows the gains to $\overline{\text{SKL}}_\text{year}$ from better performing sources and detectors. Comparing Systems B and C shows a crossover in their $\overline{\text{SKL}}_\text{year}$ at around 32~GB, highlighting an important tradeoff between the operational performance of sources and receiver for fixed memory buffers. Namely, SatQKD systems operating with constrained memory buffers should focus on improving sources (minimising $\text{QBER}_\text{I}$, System C). This is because small memory buffers can only support a short signal transmission time around the maximum elevation of a satellite's trajectory, where losses are minimised. Improving the performance of the source leads to a direct improvement of $\overline{\text{SKL}}_\text{year}$. Conversely, SatQKD systems not constrained with memory buffers have a larger operational footprint that maximises the number of overpasses that generate non-zero finite keys. Improving the key generation of these systems can be supported through improved receivers with reduce $p_\text{ec}$ (System B).

We note that a higher source rate, $f_\text{s}$, can improve the satellite overpass opportunities that generate a non-zero finite key and reduce the required memory storage. For the number of transmitted signals enabled by a limited memory buffer, a higher rate allows signal transmission over a shorter time window around $\theta_\text{max}$, where the satellite-OGS range is at its smallest, corresponding to a lower average loss. This improves both the received block length and the overall error rate. Also, the minimum amount of buffer required to generate the secret key is reduced due to more efficient transmission during the lower loss segment of an overpass. To illustrate this, consider a zenith pass with time-window of 444~seconds and a source with repetition rate of 100~MHz, which requires 22.2~GB of random bits.  If the repetition rate is increased to 500~MHz, then the same data can be transmitted in 88.8~seconds, five times less. One can thus focus the transmission at higher elevation angles, which have less loss and lower errors. The raw data for the 500~MHz source leads to a greater amount of secret key. It follows that a 500MHz source could generate the same key length as a 100~MHz source, using fewer pulses and therefore fewer random bits.

%


\subsection{Source intensity uncertainties}
\label{sec:int_fluc}

\noindent
Standard analyses of WCP decoy-state BB84 protocols usually assume perfect device operation leading to idealised key rates with optimised intensities. We can consider various deviations from ideality, such as a source with fixed and known intensities operational during the entire integration time of a satellite overpass. Active stabilisation of pulse intensities by continuous monitoring and feedback is possible~\cite{Lucamarini2013_OE} but may be limited by inherent power monitor measurement uncertainties. Instead, instantaneous offsets and long-term drifts in the intensity values lead to parameter uncertainties that are an important departure from the fixed operating intensity assumption, which directly impacts the security of distilled finite keys for two reasons. First, source intensity uncertainties can be exploited in general attacks~\cite{Yoshino2018_npjQI} which may be exacerbated in SatQKD with small block sizes. Second, the estimated vacuum and single-photon yields will differ significantly from true expectation values, potentially leading to an underestimation of the required privacy amplification to ensure security.

Several recent works have looked at this general problem by accounting for the uncertainties in source intensities directly within the parameter estimation~\cite{Wang2007_PRA, Wang2008_PRA, Hu2010_PRA, Wang2016_PRA,Liao2017_N}. This changes the estimates of the quantities that appear in Eq.~(\ref{eqn:skl_lim_result}) and could also change the secret key formula itself. A different scenario has also been considered~\cite{airqkd2022} where the existing formalism described in Refs.~\cite{Sidhu2022_npjQI,Lim2014_PRA} is used, but where one assumes that the true intensities are uncertain, though not necessarily fluctuating during a transmission block. This uncertainty results either from measurement uncertainties in the power monitors or from drifts in the calibration settings. We note that in~\cite{airqkd2022} the channels did not vary in time during a transmission block, in contrast to the SatQKD case that we consider here.

\begin{figure}[t!]
    \centering
    \includegraphics[width=\columnwidth]{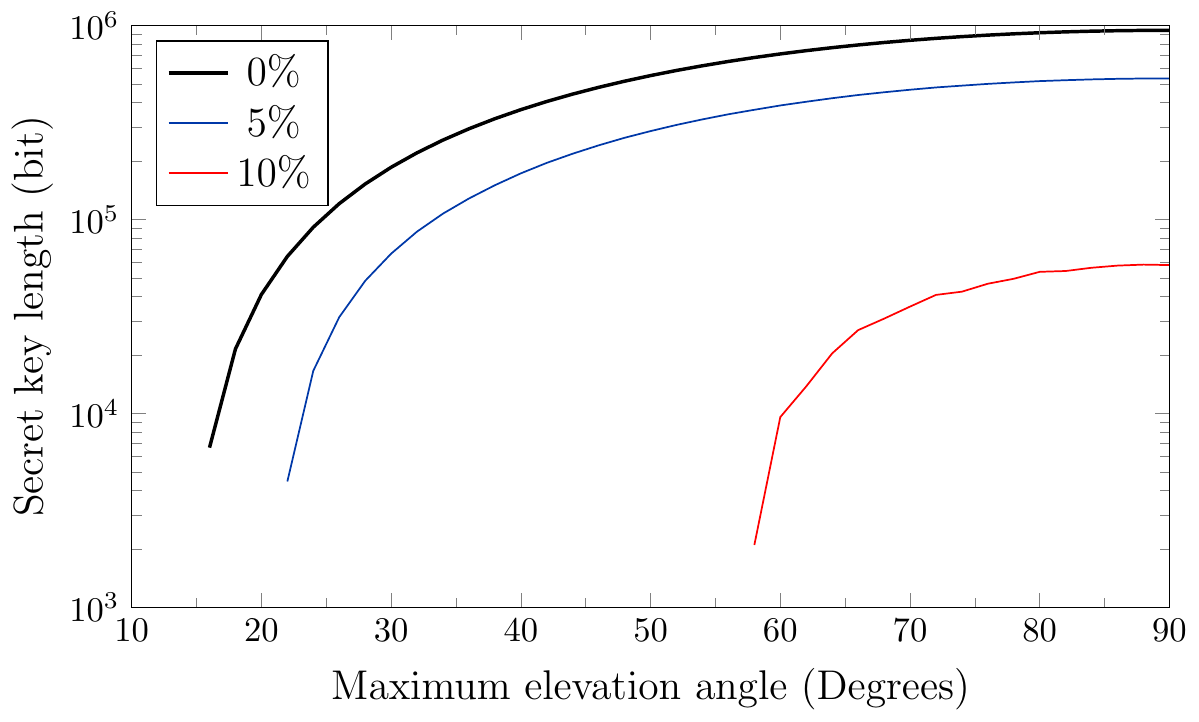}
    \caption{\textbf{Impact of source intensity uncertainty}. The signal and decoy state intensity values may independently deviate from their assumed values $\mu_j$ by fraction $f$. The per-pass SKL taking into account these intensity uncertainties for $f = 0\%,\ 5\%,\ 10\%$ are shown for different overpass geometries.}
    \label{fig:unstable_sources}
\end{figure}%

In this work, as in~\cite{airqkd2022}, we model the impact on the SKL of uncertainties in the source intensities, where we have an upper bound to on the possible deviations of $\mu_j$ from the assumed/measured values. Our approach models the case where the fixed intensity values have a constant and unknown offset from their intended values. The intensities can vary from the intended values by a maximum fraction $f$ of the intended values during an overpass. The probability of the intensity values exceeding the range defined by $f$ must be less than the advertised probability of the protocol being insecure, which is determined by $\epsilon_\text{s}$. These uncertainties are considered separately for the signal and decoy states $\mu_1$ and $\mu_2$ respectively, but not for the vacuum state, since any deviations in the vacuum state due to extraneous counts have already been considered. Crucially, we consider independent uncertainties for $\mu_1$ and $\mu_2$ for all four encoded bit values.  This is a more pessimistic approach than in related works, such as \cite{Wang2016_PRA}, where it is assumed that the uncertainties for $\mu_j$ are the same for each bit value and basis.  Each intensity value is then sampled independently in the range $\mu_j \pm f\mu_j$ to determine each signal state. Since the true intensity values are unknown to Bob, we take the worst-case combination of deviations that reduces the SKL as a conservative estimate while ensuring security. The range $\mu_j \pm f\mu_j$ is sampled using different numbers of points, though it was found that only 3 points were sufficient to find the worst-case SKL. Fig.~\ref{fig:unstable_sources} illustrates the SKL as a function of $\theta_\text{max}$ for at most a 5\% and 10\% uncertainty in the source intensities. To quantify this reduction, a 5\% and 10\% uncertainty in the source intensities reduces the annual SKL by a significant factor of 2 and 43 respectively. From this reduction, it is clear that source intensity uncertainties have a profound impact on the attainable SKL that significantly reduces the SatQKD operational footprint. For large uncertainties, it is therefore likely that the SKL will be zero for many of the satellite overpass opportunities. This highlights the importance of including the effects of uncertainties in the description of the power monitors. Active stabilisation of intensities in conjunction with high-accuracy power monitoring is important to allow operation close to the desired performance.


\section{Discussion}
\label{sec:conc}

\noindent
Existing analyses of satellite-based QKD (SatQKD) assume an ideal, fully optimised parameter space to determine the maximum finite key rate. In practice, it is difficult to engineer the control of each parameter for different satellite overpasses. Therefore, these analyses effectively serve as an upper bound to the expected performance of SatQKD. We show that SatQKD operates with limited operating margins. It is therefore of immediate practical relevance to investigate the performance of SatQKD with a reduced parameter space optimisation to reflect restrictions on system operations and deployment, and to understand its robustness to additional losses and system imperfections. Further, the limited volumetric space, weight, and power (SWaP) available on small satellites provide limited physical resources that further depart from the ideal scenario of a fully optimised parameter space. We fill this gap by establishing practical SatQKD performance limits that reflect the nature of current engineering efforts and evaluate the impacts of limited resources on the long-term finite secret key length (SKL) generation capacity. 

First, we model the impact of a fixed receiver basis bias $P_\mathsf{X}^B$ and pulse intensities $\mu_j$ on the SKL given the impracticality of their dynamic control during transmission. The SKL can be enhanced through carefully selecting the operating values of the fixed parameters. We develop a natural approach to determining the ideal fixed parameter values, based on maximising the expected annual SKL, which can be readily generalised to any parameter set. For the nominal system specifications denoted in Table~\ref{tab:system_parameters}, this leads to the fixed parameter set $\{P_\mathsf{X}^B, \mu_1, \mu_2\}$ $=$ $\{0.84, 0.71, 0.14\}$, corresponding to the receiver beamsplitter basis bias, and signal and decoy state intensities. Despite these fixed values, we find it is possible to generate near-optimal SKLs across a wide range of overpass maximum elevation angles. While larger $P_\mathsf{X}^B$ can generate larger SKL at high elevations, it does so at the expense of zero secret key at lower elevations due to worse parameter estimation. SatQKD missions should therefore carefully choose the fixed OGS bias to address the tradeoff between a maximised single-pass SKL and the long-term key generation capacity. Our optimal fixed value of $P_\mathsf{X}^B = 0.84$ balances this tradeoff to achieve close to optimal performance with fixed intensities. The optimum set of $\{P_\mathsf{X}^B, \mu_1, \mu_2\}$ will require re-evaluation for different SatQKD systems, especially in a large-scale network with several OGSs and a heterogenous space segment. Further trade-offs will have to be considered to establish a set of standard system parameters based on operational and application-specific factors.

Next, we illustrate the significant impact of limited QRNG resources that drive the source on the expected annual SKL. For the nominal system, increasing the memory buffer from 8~GB to 32~GB substantially increases the expected total annual SKL from 0.81~Gb to 3.94~Gb, corresponding to $3.16 \times 10^{6}$ and $1.54 \times 10^{7}$ AES-256 encryption keys respectively, though there are diminishing returns for larger buffers. This insight has significant implications for design trade-offs. We provide the minimum memory buffer required to yield non-zero finite keys for different overpass geometries, providing an important benchmark to support the design of upcoming SatQKD missions. For missions with higher altitudes and source rates, the QRNG subsystem for prepare-and-measure protocols will be increasingly crucial for sustained operations. High-speed QRNGs with sufficient rate for real-time driving of the source, together with ring-buffers and real-time reconciliation would obviate the need for extremely large random number stores, but will have further system design implications for SWaP-C and required communications capabilities.

Finally, we investigate the impact of uncertainties in the signal and decoy state intensities on the SKL. Maintaining fixed intensity values require perfect sources during the entire integration time of a satellite overpass. In practice, imperfect knowledge of the transmitted state intensities directly impact the security and amount of distilled finite keys whilst maintaining security. We find that these uncertainties have a profound impact on the SKL and highlight the importance of the accuracy of power monitors. Actively stabilising the intensities close to their intended values is also important to approach the optimal performances as modelled.

This study opens up a number of interesting open problems that would extend the scope and applicability of this work. First, a more comprehensive quantum channel model that includes elevation and azimuthal-dependent background light distributions, cloud cover, seasonal weather effects, and other location-dependent effects would provide a more representative performance analysis for detailed OGS siting studies. Second, different orbits and altitudes could also be modelled, the optimum altitude to maximise the integrated key generation footprint, hence its expected annual SKL, could be derived in particular. Third, implementing error correction and privacy amplification can be demanding for SatQKD. While these steps do not have to occur during the quantum transmission phase (the limited overpass time and quantum optical channel is the main bottleneck we consider in this work), modelling any inefficiencies would warrant an analysis in its own right. In particular, exploring the impact of limited resources to efficiently implement and measure error syndromes could impact the security and correctness of finite keys. Finally, an interesting extension toward the aim of establishing a global quantum network would be in exploring additional cost and performance trade-offs to reveal deeper insights into performance bottlenecks in SatQKD.

\appendix

\section*{Methods}
\label{sec:methods}

\subsection{Loss modelling}
\label{subsection:loss_modelling}

\noindent
In this section, we introduce the notation and the underlying loss model. In particular, we provide details on our model for the elevation and wavelength-dependent losses for any satellite overpass geometry. Recall that to determine the finite key, we need to determine the expected detector count statistics as a function of time and the operational source wavelength $\lambda$. Therefore, we first determine the instantaneous link efficiency as a function of elevation $\theta(t)$, range $R(t)$, and source wavelength $\lambda$, which captures all systematic and channel losses. Our method to determine the link efficiency differs from our approach in Ref.~\cite{Sidhu2022_npjQI} where we used empirical results published by Micius. In this work, we use a more physically motivated approach that will allow greater flexibility in the analysis and applications that can be considered, such as the effects of OGS positioning. Despite this change, the results of the two methods closely match for elevations above $10^\circ$ which provides confidence in the new approach.

We write the link efficiency as
\begin{align}
  \eta_{\lambda}\left(\theta\right) = \eta_\text{diff}\left(\lambda,\theta\right) + \eta_\text{atm}\left(\lambda,\theta\right) + \eta_\text{int},
\label{eqn:tot_loss}  
\end{align}
in units of decibels (dB) and where we have three distinct loss contributors. The first term $\eta_\text{diff}$ defines losses from diffraction effects, $\eta_\text{atm}$ from atmosphere effects that include scattering and absorption, and $\eta_\text{int}$ defines a fixed elevation-independent intrinsic system efficiency corresponding to internal losses, and beam misalignment. Eq.~\eqref{eqn:tot_loss} provides a general approach to modelling losses for any SatQKD system. Once a satellite overpass trajectory is defined, we use Eq.~\eqref{eqn:tot_loss} to determine the loss for every second of the overpass to estimate the total count statistics. A single block is then constructed from the entire overpass data, and finite statistics incorporated to maintain composable security. Details for each loss contributor are provided below.

\begin{figure}[t!]
    \centering
    \includegraphics[width=\columnwidth]{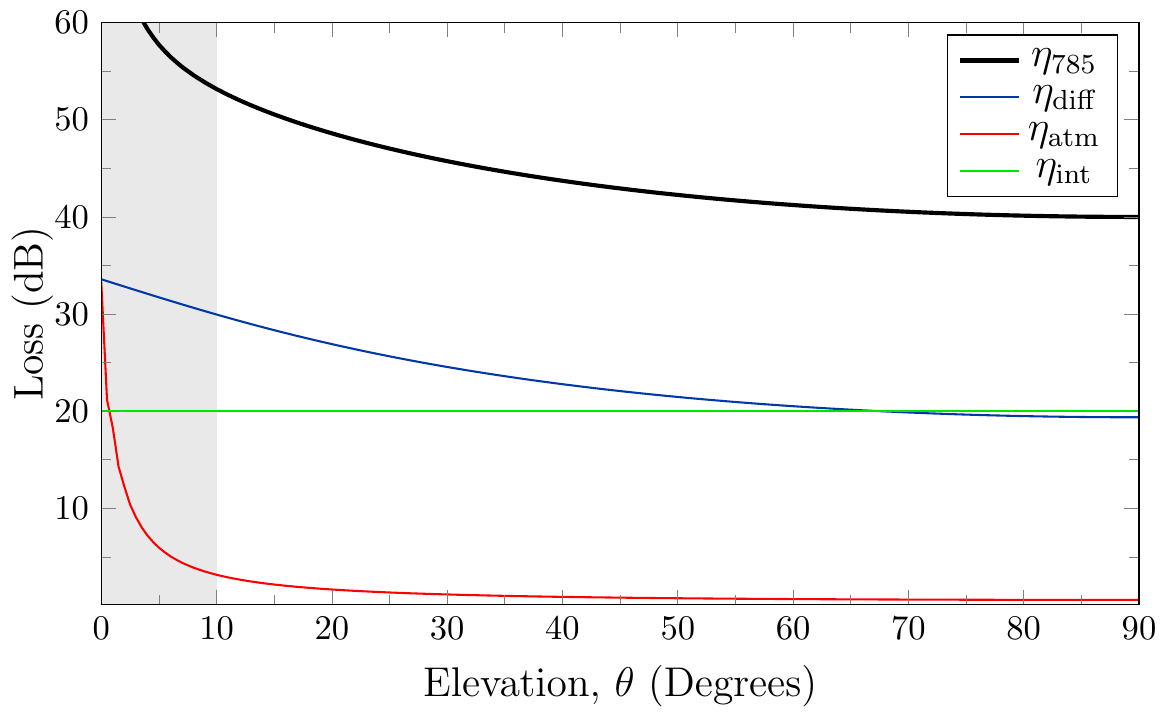}
    \caption{\textbf{Link efficiency as a function of elevation}. Each contributor to the total loss is illustrated for $\lambda=785$~nm. Both diffraction and atmospheric losses vary with elevation and increase with decreasing elevations. The solid black line illustrates the total link efficiency. The loss axis is truncated at 60 dB, with the worst link efficiency being $\eta_{785}=87$~dB at $0^\circ$. The loss values in the gray region, where the elevation falls below 10$^\circ$ are not used in the key length simulations.}
    \label{fig:loss_vs_elevation}
\end{figure}%
%


\subsubsection{Diffraction losses}
\label{subsubsec:diff_loss}

\noindent
A dominant contribution to loss is diffraction, which broadens the beam after the signal propagates through the satellite's transmitter aperture, $T_\mathsf{X}$. The amount of beam broadening depends on a number of factors, including the channel range $R(t)$, $T_\mathsf{X}$, and the source wavelength $\lambda$. Here, we take a standard approach to estimate diffraction losses by calculating the far-field Fraunhofer diffraction of a initial truncated Gaussian field distribution with a beam waist of $w_0$ at the transmission aperture. We calculate the probability that a single photon exiting the transmit aperture is collected by the receiver aperture from the ratio of the integrated power density across the transmitter aperture, $P_T$, and the receiver aperture, $P_R$,
\begin{align}
  \eta_\text{diff}\left(\lambda,\theta\right) = -10 \log_{10}\left(\frac{P_R}{P_T}\right).
\end{align}
Since we are using a weak coherent pulse (WCP), there is no optimal beam waist provided there is no constraint on beam power~\cite{Bourgoin:2013fk}. For a downlink configuration with a WCP source, it is optimal to have the beam waist be as large as possible to achieve close to ideal far-field diffraction. However, practical constraints on the source power will impose a limit to flatness of the Gaussian across the transmission aperture. Therefore, we set the beam waist to be in the order of the transmitter aperture diameter, $w_0 = T_\mathsf{X}/2$. The impact of a central beam obscuration due to secondary mirrors typical of Cassegrain-type reflecting telescopes could be considered~\cite{Bourgoin:2013fk} but has no significant impact on the analysis.


\subsubsection{Atmospheric attenuation}
\label{subsubsec:atm_loss}

\noindent
The second contributor to the instantaneous link efficiency arises from atmospheric attenuation from absorption and scattering from molecules and particulate matter. The magnitude of these atmospheric losses depends on the wavelength and the satellite's elevation, which determines the length of the quantum channel through the atmosphere. We use MODTRAN to model atmospheric propagation and determine the transmissivity, $T_{\lambda}(\theta)$, for a given wavelength as a function of elevation. MODTRAN is a software that solves the radiative transfer equation to provide a standard atmospheric band model~\cite{Modtran_inproceedings}.

The atmospheric loss contribution is then calculated from the transmissivity,
\begin{align}
  \eta_\text{atm}\left(\lambda,\theta\right) = -10 \log_{10}\left(T_{\lambda} (\theta)\right),
\end{align}
where the wavelength and elevation dependence is made clear.


\subsubsection{`Intrinsic' system loss}
\label{subsubsec:intrinsic_loss}

\noindent
The final loss contributor is denoted the `intrinsic' system loss $\eta_\text{int}$ that combines several sources. We simplify the analysis by taking this to be fixed, i.e. elevation/time independent. Within our loss budget, the intrinsic system loss combines two distinct loss contributors. First, we conservatively assign a fixed loss of 12~dB to the overall electro-optical inefficiency of the OGS system, which is comprised of 3~dB each from,
\begin{enumerate}
\item photon detection efficiency Si-SPAD,
\item quantum receiver optics,
\item collection telescope,
\item interface and adaptive/tip-tilt optics between telescope and quantum receiver.
\end{enumerate}
We also lump together losses due to an imperfect, non-diffraction limited, beam (beam quality parameter $M^2 > 1$), turbulence induce beam wander and spreading, and transmitter pointing errors. For simplicity, we assign a fixed and conservative value of 8~dB to such non-ideal beam propagation induced losses. Therefore, in this work, we set
\begin{align}
  \eta_\text{int} = 20.0~\mathrm{dB},
\end{align}
which brings the total minimum loss at zenith to $\slm=40$~dB. Elevation dependence of the turbulence-induced losses has been considered in other works but is neglected for the moment in this work. More detailed modelling of turbulence and pointing losses can be found in~\cite{trinh2022statistical} and references therein. Under-estimation of these losses is compensated in part by conservative estimates made elsewhere in $\eta_\text{int}$.

Note that these are conservative estimates that may be more indicative of practical SatQKD systems. If we are able to engineer better performances and achieve highly optimised operation, then we can further reduce the receiver and transmitter apertures for increased portability, while maintaining the values of $\slm$ analysed here. These losses are consistent with the recent mobile OGS designed for the Micius mission~\cite{Ren2022arxiv}.


\subsection{Error correction for one-way information reconciliation}
\label{subsec:error_corr_term}

\noindent
An important step for any QKD protocol is error correction, which identifies and corrects errors due to vacuum events and transmission errors. For this step, Alice and Bob publicly announce $\lambda_\text{EC}$ bits that are assumed known to Eve through a round of classical communication. The number of bits $\lambda_\text{EC}$ depends on the error rate, which is a practical implementation we estimate during the parameter estimation stage. For our simulation, we use an estimate of $\lambda_\text{EC}$ that varies with the quantum bit error rate (QBER), $Q$, and the data block size, $n_\mathsf{X}$. A common approach to modelling the number of error correction bits required during information reconciliation is through $f_\text{EC} n_\mathsf{X} h(Q)$, where $f_\text{EC}$ is the reconciliation factor efficiency and we recall that $h(x)$ is the binary entropy function. The value for $f_\text{EC}$ is crucially larger than unity, and often chosen within the range 1.05 to 1.2, to account for inefficiencies in the error correction protocol. While this approach is well-suited to determining the optimal secret key length, it is assumed that the reconciliation factor efficiency is independent of $Q$, $n_\mathsf{X}$, and the required correctness $\epsilon_\text{c}$. Since SatQKD operates within the finite-key regime, these parameters can vary significantly, however. An improved estimate of the reconciliation factor efficiency would enable a higher SKL under finite statistics.

\begin{figure}
    \centering
    \includegraphics[width=\linewidth]{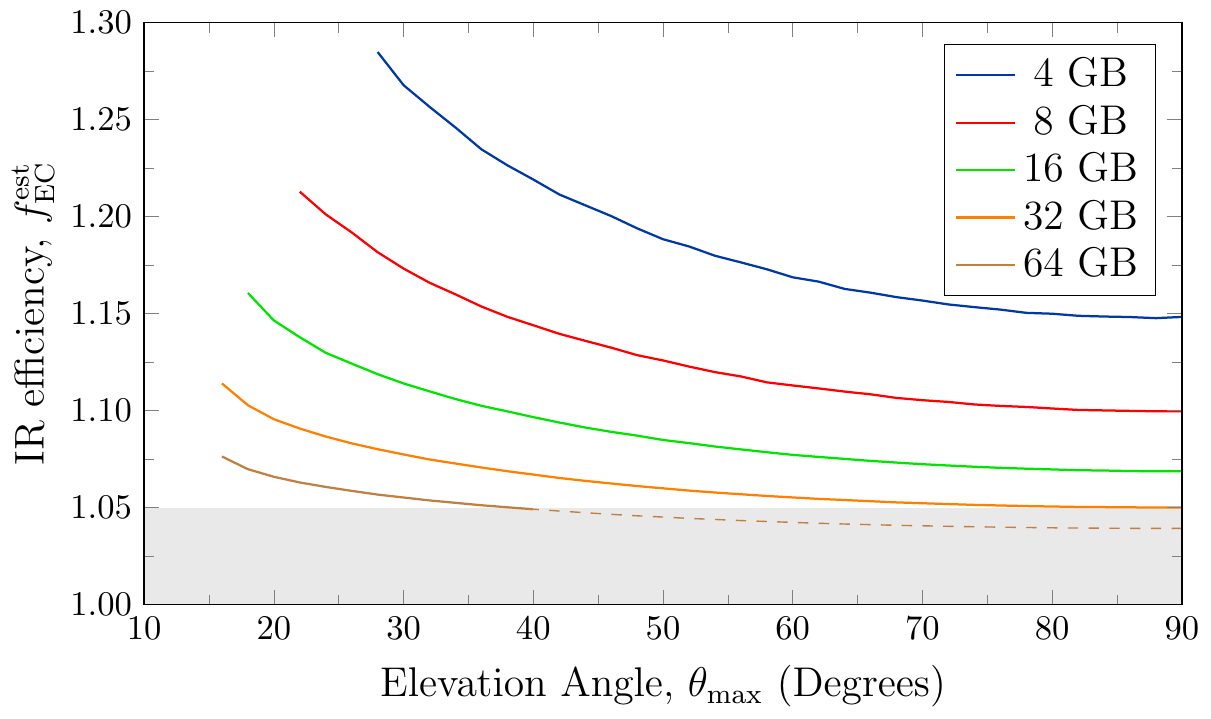}
    \caption{\textbf{One-way information reconciliation efficiency.} We estimate $\bm{f_\text{EC}^\text{est}}$ as a function of satellite overpasses with maximum elevation angle $\theta_\text{max}$ for different memory buffers $m_b$. For data representative of current engineering efforts, $f_\text{EC}^\text{est}$ remains larger than 1.05, which is the lowest quoted achievable efficiency in the literature and is illustrated by the gray region corresponding to optimistic efficiencies.}
    \label{fig:IR_eff}
\end{figure}

The amount of information leaked to the eavesdropper during information reconciliation is usually impossible to determine exactly. Therefore it is often upper bounded by $\log\abs{\mathcal{M}}$, where $\mathcal{M}$ denotes the error syndrome. For one-way reconciliation, the size of this error syndrome (in bits) has the following tight lower bound~\cite{Tomamichel2017_QIP}
\begin{align}
\begin{split}
    \lambda_\text{EC} = & \; n_{\mathsf{X}} h(Q) + n_{\mathsf{X}} (1 - Q)\log\left[\frac{(1 - Q)}{Q}\right]\\
    &- 
    \left(F^{-1}( \epsilon_\text{c};n_{\mathsf{X}},1-Q,) - 1\right) \log\left[\frac{(1 - Q)}{Q}\right]\\
    &- \frac{1}{2} \log(n_{\mathsf{X}}) - \log(1/\epsilon_\text{c}),
\end{split}
\label{eq:lambdaec}
\end{align}
where $F^{-1}$ is the inverse of the cumulative distribution function of the binomial distribution. We use this estimate for the number of error correction bits to determine the optimised SKL. We note that for large block sizes 
\begin{align}
\lim_{n_{\mathsf{X}} \rightarrow \infty} \frac{\lambda_\text{EC}}{n_{\mathsf{X}} } = h(Q),
\label{eqn:err_corr_limit}
\end{align}
such that $\lambda_\text{EC}^\infty = n_{\mathsf{X}} h(Q)$, which is the minimum possible bits allowed by information theory. This suggests that the information reconciliation (IR) factor efficiency tends towards unity $f_\text{EC}=1$, which is optimistic even for optimised low-density parity-check (LDPC) codes that can achieve high reconciliation efficiencies and require few rounds of communications~\cite{Elkouss2009}. For application in SatQKD, the IR efficiency does not approach this asymptotic limit over QBERs and data block sizes typical of realistic operation. To demonstrate this, we investigate how the IR efficiency estimate varies for the different memory buffers considered in Section~\ref{sec:mem_buffer}. Specifically, the finite-size estimate for the IR efficiency provided by Eq.~\eqref{eq:lambdaec} can be determined from the ratio $f_\text{EC}^\text{est} = \lambda_\text{EC} /  n_{\mathsf{X}} h(Q)$. Fig.~\ref{fig:IR_eff} illustrates this ratio as a function of satellite overpasses with maximum elevation angle $\theta_\text{max}$ for different memory buffers $m_b$. Note that the data block sizes increase with an increasing memory buffer, leading to better $f_\text{EC}^\text{est}$ that approaches unity. We observe that the estimated efficiency dips below the lower quoted value of 1.05 in the literature~\cite{Tomamichel2017_QIP}, which is indicated by the gray region. Recall from section~\ref{sec:mem_buffer}, that a memory buffer of 64~GB achieves near-optimal performance corresponding to the highly optimised scenario. Therefore, the correction estimate in Eq.~\eqref{eqn:err_corr_limit} does not approach the asymptotic limit of unit efficiency for SatQKD data representative of current engineering efforts and capabilities and is well suited to explore the engineering constraints that are the focus in this work.

Before concluding, we make two observations. First, a simple remedy to the error correction estimate that would hold for any data block size would be to switch to an updated model whenever the reconciliation efficiency estimated by Eq.~\eqref{eqn:err_corr_limit} falls below 1.05. That is, we can estimate the number of error correction bits required from 
\begin{align}
\lambda_\text{EC}^{\text{new}} = f_\text{EC} n_{\mathsf{X}} h(Q) \, ,
\label{eqn:err_corr_improved}
\end{align}
where $f_\text{EC}$ takes values that reflect achievable efficiencies, whenever $\smash{\lambda_\text{EC} < 1.05 n_{\mathsf{X}} h(Q)}$. Second, here we do not consider bi-directional error correction information reconciliation for SatQKD such as CASCADE~\cite{brassard1994secret}. Although it may lead to improved reconciliation efficiencies, the complexity of classical communication protocols and operations, and demands for on-board data processing are significantly greater. Hence, it may be more practical to implement one-way IR in SatQKD to simplify operations and reduce system cost and complexity using schemes such as low-density parity check (LDPC) codes~\cite{johnson2015analysis}.


\subsection{General approach optimisation of fixed parameter values}
\label{subsection:param_fixing}

\noindent
The fully optimised finite SKL is difficult to achieve since it requires active control of the entire parameter space, which may be difficult to engineer. In section~\ref{sec:param_fixing} we explored the impact of fixing the receiver basis bias $P^B_\mathsf{X}$, and the two intensity values $\mu_1$ and $\mu_2$ that are particularly challenging to change. This naturally raises the question \emph{what fixed values should a SatQKD system implement}? Here, we outline a general method to determine fixed values for the set $\mathcal{F} \in \{P^B_\mathsf{X}, \mu_1, \mu_2\}$. 

\begin{figure}[t!]
    \centering
    \includegraphics[width=\linewidth]{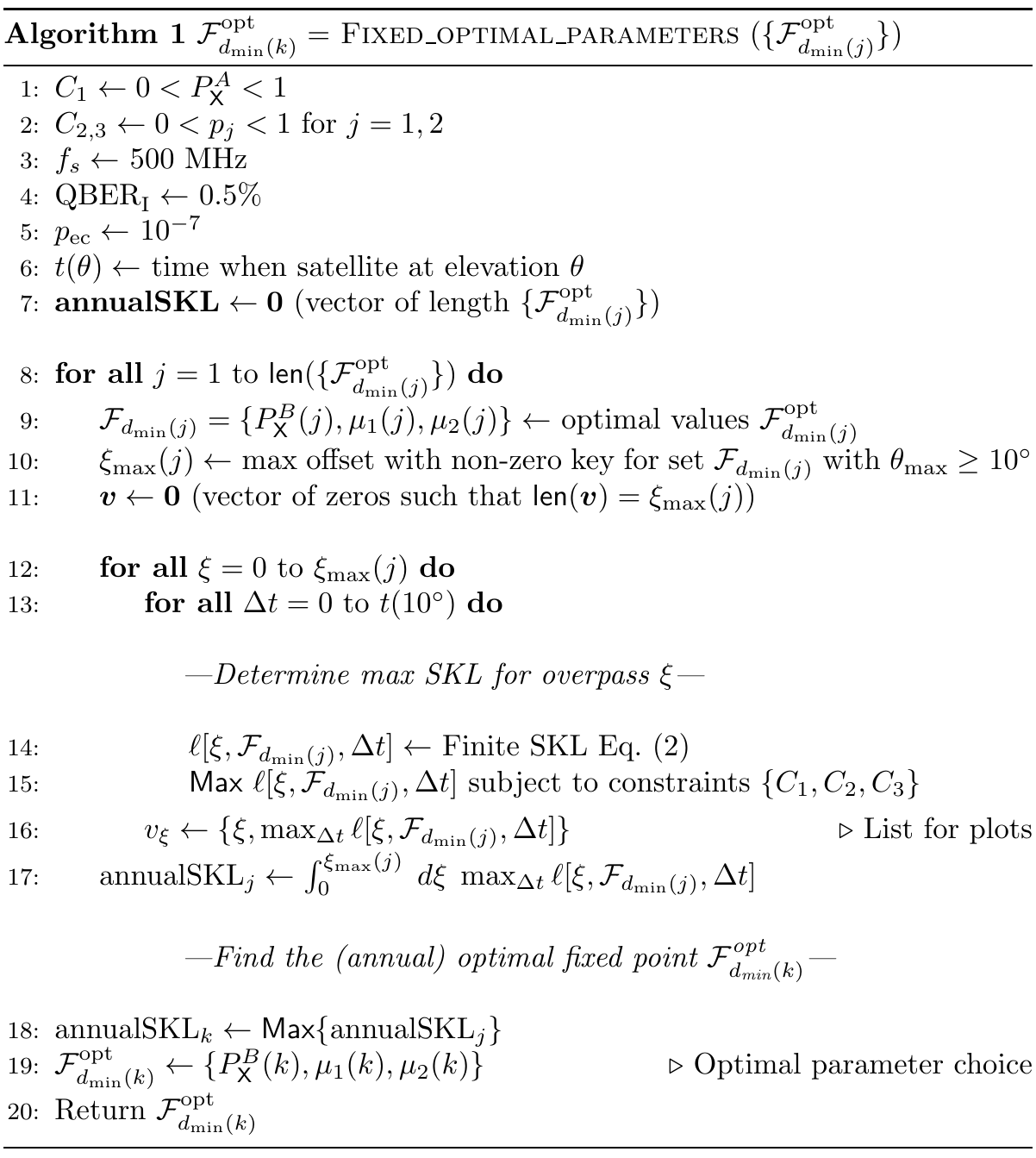}
    \caption{\textbf{Pseudocode to determine the ideal fixed parameter set}. We denote $\smash{\mathcal{F}_{d_\text{min}(k)}^\text{opt} = \{P^B_\mathsf{X}(k), \mu_1(k), \mu_2(k)\}}$ as that which maximises the performance of a SatQKD system through the expected annual SKL, which is determined from the parameter set $\mathcal{F}_{d_\text{min}(j)}^\text{opt}$ that are sampled from the fully optimised SKL vs $d_\text{min}$ plot. The list $\nu_\xi$ is used to generate the plots in this work. This algorithm can be generalised to determining the ideal values for any fixed parameter set.} 
    \label{fig:param_fix_algo}
\end{figure}

Our method follows from maximising $\overline{\text{SKL}}_\text{year}$, which is proportional to the integrated area under the SKL vs ground track distance curves, $\text{SKL}_\text{int}$~\cite{Sidhu2022_npjQI}. We first establish the fully optimised SKL as a function of $d_\text{min}$, corresponding to optimising the full parameter space. For each point $j$ along the optimised curve, we extract the set, $\mathcal{F}_{d_\text{min}(j)}^\text{opt}$, of the optimal values for $P^B_\mathsf{X}$, $\mu_1$, and $\mu_2$ for $d_\text{min}(j)$ (in units of $10^6$ m). Now fixing $\mathcal{F}_{d_\text{min}(j)}^\text{opt}$, we optimise the SKL over the remaining parameter space to determine the SKL as a function $d_\text{min}(j)$, hence $\text{SKL}_\text{int}$. This procedure is repeated for each optimised point $j$. We then choose the fixed set $\smash{\mathcal{F}_{d_\text{min}(k)}^\text{opt}}$ that maximises $\text{SKL}_\text{int}$ as the best compromise of fixed parameters. This procedure is summarised in Fig.~\ref{fig:param_fix_algo}.

\begin{figure*}
    \centering
    \includegraphics[width=\linewidth]{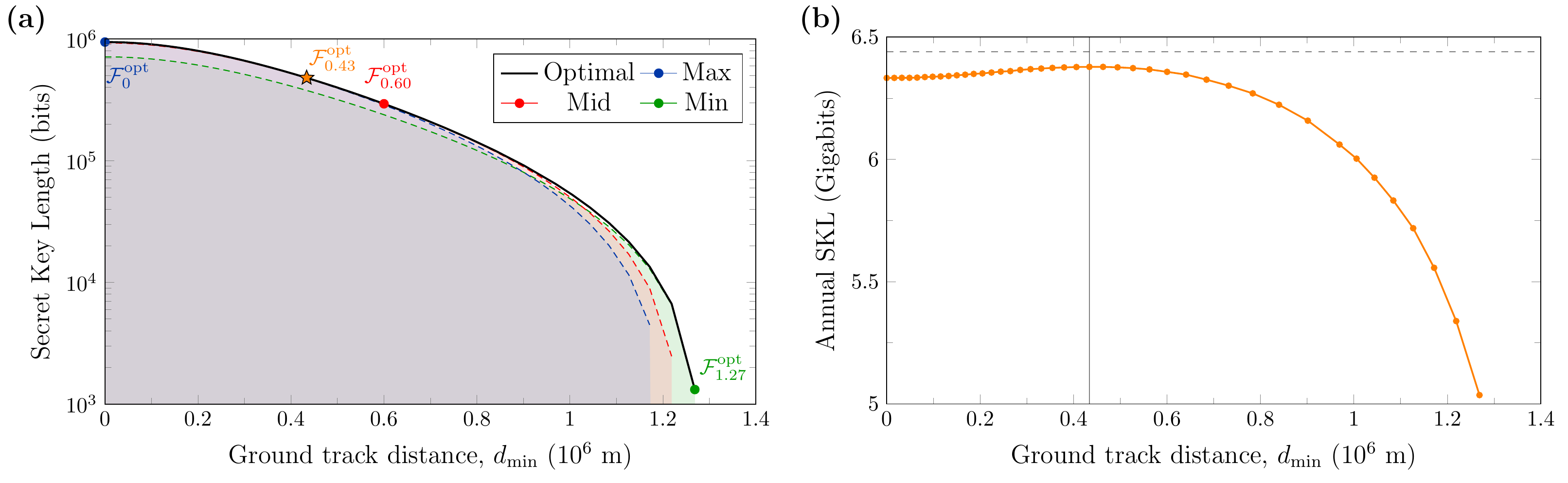}
    \caption{\textbf{SKL  
    vs $\smash{d_\text{min}}$ for fixed $\smash{\bm{\mathcal{F}}}$}. (a) The fully optimised SKL is illustrated in black, with each fixed point $j$ along the optimal curve generating the set $\smash{\mathcal{F}_{d_\text{min}(j)}^\text{opt}}$, corresponding to the optimal fixed parameter values at ground track distance $d_\text{min}(j)$ (in units of $10^6$~m). The SKL for three illustrative fixed sets, $\smash{\mathcal{F}_0^\text{opt}}$, $\smash{\mathcal{F}_{0.60}^\text{opt}}$, and $\smash{\mathcal{F}_{1.27}^\text{opt}}$, are optimised over the remaining parameter space with their corresponding areas shaded to determine the expected annual SKL. The ideal fixed data set is highlighted with an orange star at $d_\text{min} = 0.43 \times 10^6$~m. (b) Variation in the expected annual SKL for each fixed set $\smash{\mathcal{F}_{d_\text{min}(j)}^\text{opt}}$. The vertical solid line corresponds to the parameter set that maximises the estimated annual SKL and the horizontal dashed line to the annual SKL with no constraints.} 
    \label{fig:param_fix_choice}
\end{figure*}

Fig.~\ref{fig:param_fix_choice} illustrates this procedure for choosing the ideal fixed set $\smash{\mathcal{F}_{d_\text{min}(k)}^\text{opt}}$ that optimises $\overline{\text{SKL}}_\text{year}$. In Fig.~\ref{fig:param_fix_choice}(a), the optimal SKL is illustrated in black. Three illustrative fixed sets $\mathcal{F}_{d_\text{min}(j)}$ are sampled to correspond to the maximum, median, and minimum non-zero SKLs values and are shown in dashed blue, dashed red, and dashed green respectively. We first note that fixing the values for $\mathcal{F}$ has little impact on the SKL over the entire range of satellite overpass trajectories. This reassuringly demonstrates that SatQKD systems operating with a fixed subset of parameters $\mathcal{F}$ do not lead to a large departure from the optimal performance with only a small observed impact on the SKL generation performance. Second, it is possible to improve the SKL by carefully choosing the fixed values for $\mathcal{F}$. The ground track distanced furthest away from the sampled point $j$ along the optimal curve deviates most from the optimal performance. This suggests that the fixed parameter set should be chosen closer to the centre of the curve, since this would maximise the robustness of the SatQKD systems to the widest variety of satellite overpasses leading to the largest annual expected SKL. This specific dependence on the fixed parameter set and the annual SKL is illustrated in Fig.~\ref{fig:param_fix_choice}(b). The peak annual SKL corresponds to the ideal fixed set $\smash{\mathcal{F}_{0.43}^\text{opt} = \{0.841, 0.709, 0.139\}}$. This establishes the fixed values used in section~\ref{sec:param_fixing}. Our method is general and can be extended to determining the ideal values for any alternative subset of fixed parameter sets. Finally, we reassuringly find that despite the constrained parameter space, the estimated annual SKL with these fixed parameters is close to the fully optimised case, shown with the dashed horizontal line in (b).

We note that there is the possibility that a greater $\overline{\text{SKL}}_\text{year}$ could be achieved with a parameter set outside of the per-pass optima but as the presented procedure closely approaches the upper bound, a search for such values may not be worthwhile.

%


\section*{Data availability}

\noindent
The raw output files from the simulations used to generate data in this work are available upon reasonable request. All material requests should be made to J.S.S. 


\section*{Code availability}

\noindent
The SatQuMA v1.1 simulation Python suite is available at Ref.~\cite{Sidhu2021arxiv}. Modified code used to generate all results in this work is accessible on GitHub \href{https://github.com/cnqo-qcomms/SatQuMA/}{https://github.com/cnqo-qcomms/SatQuMA/}. It implements a minor modification of SatQuMA v1.1 to handle fixed parameters that have currently not been released as a stand-alone package. 


\bibliographystyle{ieeetr}


\section*{Acknowledgements}

\noindent
We acknowledge support from the UK NQTP and the EPSRC Quantum Technology Hub in Quantum Communications (grant: EP/T001011/1), and the EPSRC International Network in Space Quantum Technologies (grant: EP/W027011/1). We also acknowledge support from the UK Space Agency (NSTP3-FT-063, NSTP3-FT2-065, NSIP ROKS Payload Flight Model), the Innovate UK project ReFQ (Project number: 78161), Innovate UK project AirQKD (Project number: 45364), the Innovate UK project ViSatQT (Project number: 43037), EU QTSPACE (COST CA15220), and the EPSRC Research Excellence Award (REA) Studentship.


\section*{Author contributions}

\noindent
J.S.S. conceptualised the main ideas together with D.K.L.O., steered the direction of research, and wrote the initial draft. J.S.S., T.B., and D.M. wrote the initial version of the code (SatQuMA v1.1) that is openly available, with modifications made by J.S.S. and T.B. to obtain numerical results presented in this work. R.G.P. conducted background literature reviews. D.K.L.O. obtained funding and initiated the research. All authors contributed to selecting relevant literature, proofreading, and editing the manuscript.


\section*{Competing financial interests}

\noindent
The authors declare no competing financial interests.

\end{document}